\documentclass[preprint]{aastex}
\usepackage{graphicx}
\newcommand{\beq}{\begin{equation}}
\newcommand{\eeq}{\end{equation}}
\newcommand{\bea}{\begin{eqnarray}}
\newcommand{\eea}{\end{eqnarray}}
\newcommand{\bctr}{\begin{center}}
\newcommand{\ectr}{\end{center}}

\usepackage{natbib}

\begin{document}
\title{Ideal bandpasses for type Ia supernova cosmology}

\author{Tamara M. Davis\altaffilmark{1}, Brian P. Schmidt}
\affil{Australian National University, via Cotter Road, Weston Creek, ACT, 2611, AUSTRALIA}
\affil{tamarad@mso.anu.edu.au, brian@mso.anu.edu.au}
\and
\author{Alex G. Kim}
\affil{Lawrence Berkeley National Laboratory, 1 Cyclotron Road, Berkeley, CA 94720, USA}
\affil{agkim@lbl.gov}
\altaffiltext{1}{Lawrence Berkeley National Laboratory, 1 Cyclotron Road, Berkeley, CA 94720, USA}

\keywords{instrumentation: miscellaneous --- space vehicles: instruments --- supernovae: general}

\begin{abstract}
To use type Ia supernovae as standard candles for cosmology we need accurate broadband magnitudes.  In practice the observed magnitude may differ from the ideal magnitude-redshift relationship either through intrinsic inhomogeneities in the type Ia supernova population or through observational error.   Here we investigate how we can choose filter bandpasses to reduce the error caused by both these effects.  
We find that bandpasses with large integral fluxes and sloping wings are best able to minimise several sources of observational error, and are also least sensitive to intrinsic differences in type Ia supernovae.  The most important feature of a complete filter set for type Ia supernova cosmology is that each bandpass be a redshifted copy of the first.  We design practical sets of redshifted bandpasses that are matched to typical high resistivity CCD and HgCdTe infra-red detector sensitivities.  These are designed to minimise systematic error in well observed supernovae, final designs for specific missions should also consider signal-to-noise requirements and observing strategy.  In addition we calculate how accurately filters need to be calibrated in order to achieve the required photometric accuracy of future supernova cosmology experiments such as the SuperNova-Acceleration-Probe (SNAP), which is one possible realisation of the Joint Dark-Energy mission (JDEM).  We consider the effect of possible periodic miscalibrations that may arise from the construction of an interference filter.  
\notetoeditor{I have moved most of the figures to the end of the document, as requested in the instructions to authors.  The only figure I left in line with the text is the figure written in latex depicting a trapezoid (currently figure 1).  I have left this in place because it should sit at this point in the text.  In the published version this trapezoid should appear as Figure 2, while the first of the figures at the back (currently figure 2) should be restored to Figure 1 (I refer to it first.)  If it is any help I have left ``PLACEHOLDERS'' at the approximate positions in the text where figures should appear.  Regards, Tamara.}
 \end{abstract}


\section{Introduction}

Many cosmological studies using type Ia supernovae (SNe Ia) have been restricted to using only those filter sets that are available on the telescopes that make the observations.\footnote{The High-z SN team is one exception, having designed and purpose built a filter set for their SN Ia measurements.}  However, upcoming dedicated SN Ia cosmology cameras, such as those being designed for the Joint Dark-Energy Mission \citep{jdem03}, can gain considerable advantage through using a set of filters designed specifically for measuring SNe Ia.  A clever choice of bandpasses can improve our ability to constrain systematic errors and thus our ability to extract scientific information.  In this paper we study the effect of bandpass choice on SN Ia measurements and design an ideal bandpass set for supernova Ia studies.

For supernova Ia studies we need bandpasses covering the whole wavelength range appropriate for the survey depth to enable us to probe the same spectral region of highly redshifted supernovae as we do for nearby supernovae.  Due to the large redshift range being sampled we need to be able to accurately correct for the brightness of supernovae that are red-shifted away from the centre of a bandpass (K-correction), and we would prefer the required K-correction to be small in order to minimise any systematic error we may introduce in our analysis.  Accurate K-corrections are most easily achieved through a set of bandpasses that are redshifted copies of one another so the required K-corrections are periodic in redshift.  

Type Ia supernovae are a very homogeneous set, but they are not all identical~\citep{filippenko92,phillips92,leibundgut93,benetti04,benetti05,james05}.   Therefore, we also benefit  from choosing bandpasses through which SNe Ia appear most homogeneous --- that is, choosing bandpass shapes that are not overly sensitive to the spectral diversity of the type Ia supernova population.   We would also like bandpasses that give good colour information so that we can measure the effect of extinction and account for it.  The total throughput of the bandpasses is important in determining exposure times, and the greater the throughput the faster the survey can be completed.  However, we are restricted to creating bandpasses that fit in the sensitive regions of available detector technology and can be created with current (or near future) filter technology.  

\begin{figure}[t]\begin{center}
\includegraphics[width=85mm]{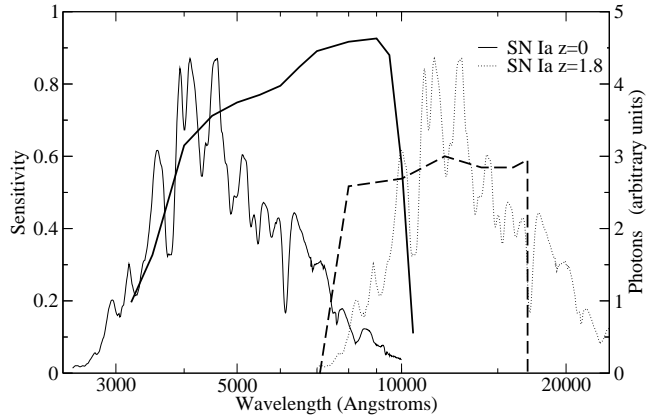}
\caption{\footnotesize Typical SN Ia spectrum at maximum light~\citep[][with 2005 update]{nugent02} superimposed on the sensitivity of a typical optical detector (high-resistivity, deep-depleted CCD, thick solid line) and the sensitivity of a typical NIR detector (HgCdTe detector with 1.7$\mu$m cutoff, thick dashed line).  The spectrum that peaks at 4000\AA ~(solid line) is for a SN Ia at zero redshift while the spectrum that peaks at 11000\AA ~(dotted line) is for a SN Ia at a redshift of 1.8.  The wavelength axis is logarithmic.}
\label{fig:SNspectra-Detectors3}
\end{center}\end{figure}

Type Ia supernova spectra at zero redshift are bright over a range of 2500\AA$-$10000\AA.  The ultra-violet (UV) cut-off is quite sharp, but the IR fall-off is gradual --- approximately the Rayleigh-Jeans tail.  The rest frame peak brightness at the time of maximum light occurs at about 4000\AA.  At earlier epochs this shifts slightly to lower wavelengths, while at later epochs the spectrum peaks at longer wavelengths.  
At a redshift of 1.8 this range translates to 7000\AA$-$28000\AA\ with a peak at 11200\AA.
A typical maximum light SN Ia spectrum is shown superimposed on typical optical and near-infra-red (NIR) detector sensitivies in Figure~\ref{fig:SNspectra-Detectors3}. 

As discussed, any set of bandpasses for SN Ia cosmology should consist of logarithmically-distributed bandpasses --- each bandpass being a redshifted copy of the first --- in order to minimise uncertainty in K-corrections (see Sect.~\ref{sect:kcorr}).  The free variables include
(1) The shape of a bandpass: transmittance as a function of wavelength.  Some shapes are less sensitive to the intrinsic differences in SNe Ia than others.  (2) The width of a bandpass.  Wider bandpasses increase throughput but may decrease colour discrimination ability.  (3) The spacing of the bandpasses.  This is highly constrained by the detector sensitivities and bandpass width.  (4) The number of bandpasses.  The survey can be completed more quickly for fewer bandpasses.  (5) The stability and knowledge of the transmission of the bandpass.  

Note that throughout this article we concentrate on the {\em bandpasses} required for supernova studies.  The filters that realise these bandpasses need to be designed while taking into account the mirror reflectivity and detector sensitivity of the telescope for which the filters are made.  Also, when designing final filters for a specific mission the signal-to-noise advantage of wider bandpasses should be considered in light of the planned observing strategy.

Many of the examples in this paper can be applied to the SuperNova Acceleration Probe (SNAP) \citep{snap04}, a proposal for the JDEM that concentrates on broadband photometry of type Ia supernovae.  This paper is also relevant to other JDEM proposals that use type Ia supernovae, even if they use spectra rather than broadband photometry, since the process of converting a spectrum to a measure of luminosity generally assumes some passband.

This paper has three main parts.  In Section~\ref{sect:filtershapes} we discuss general bandpass shape considerations, examining the benefits of sloping bandpass edges vs.\ top-hat profiles as well as bandpass width and bandpass spacing.  In Section~\ref{sect:canonical} we design several bandpass sets, tuned to typical detector sensitivities, that allow measurements of supernovae over a range of redshifts out to at least $z=1.8$.  These first two parts assume that filters to realise these bandpasses can be manufactured and calibrated perfectly and the only scatter in our measurements comes from the inhomogeneities in the SN Ia population.  In the final section we investigate the effect of filter calibration errors on SN Ia magnitude calculations.  We test both random and periodic calibration errors (periodic calibration errors may be expected when using interference filters), and identify frequencies that are particularly damaging for SN Ia measurements.

\section{Bandpass shape considerations}\label{sect:filtershapes}

Here we investigate the role that bandpass shape will play in the systematic and stochastic error of future SN Ia cosmology missions.  Possible bandpass shapes range from the `top hat' profile with very sharp cut-off, to a very curved bandpass with a smooth transition from no throughput to maximum throughput.  The bandpasses may be symmetrical or asymmetrical.

The bandpass shape can affect the quality of data collected in a variety of ways.  Diversity in the SN Ia population causes a natural scatter in the data (even after the light curve width-luminosity relation has been corrected for).  This scatter can be minimised with a clever choice of bandpasses.  Bandpasses also determine the accuracy with which we can measure colour, and thus deduce extinction correction and other colour dependent traits.  

Minimising the scatter due to SN Ia diversity is important not only for minimising the number of supernovae needed to get good statistics but also to reduce the impact of many types of systematic error, such as a bias due to population evolution.   For example if we use a model of the supernova spectrum to calculate the K-correction for each supernova, and this model spectrum is not a true representation of the mean, then we may introduce a systematic magnitude offset to the supernovae.   An evolution of the SN Ia population with redshift would create a shift in the mean which would introduce a spurious redshift dependent trend in average peak magnitude.  Any bandpass set that minimises the scatter due to intrinsic differences in the type Ia supernova population should also minimise these systematic effects.   

In order to investigate the effect of bandpass shape on SN Ia measurements we consider some idealised bandpass shapes.  Each bandpass is a symmetric trapezoid described by the width of the rising edge of the trapezoid ($a$) and the width of the flat top ($b$).  The integral flux (arbitrary units) is given by $a+b$ (this is the area under the trapezoid).  We use symmetric trapezoids only for the simplicity of reducing the number of parameters we test.  Asymmetric bandpasses follow similar trends, and may be more suitable for a final design based on the detailed matching of bandpasses to detector and mirror sensitivities.
\setlength{\unitlength}{8mm}
\begin{figure}[h!]\bctr
\begin{picture}(10,5)
	\thicklines
	\put(0,0){\line(3,4){3}}
	\put(3,4){\line(1,0){4}}
	\put(7,4){\line(3,-4){3}}
	\put(0.05,0){\vector(1,0){2.9}}
	\put(2.95,0){\vector(-1,0){2.9}}
	\put(3.05,0){\vector(1,0){3.9}}
	\put(6.95,0){\vector(-1,0){3.9}}
	\put(7.05,0){\vector(1,0){2.9}}
	\put(9.95,0){\vector(-1,0){2.9}}
	\put(8.5,0.3){\makebox(0,0){$c$}}
	\put(1.5,0.3){\makebox(0,0){$a$}}
	\put(5,0.3){\makebox(0,0){$b$}}
	\put(3,0){\line(0,1){4}}
	\put(7,0){\line(0,1){4}}
\end{picture}\ectr
\caption{\footnotesize Trapezoidal bandpass.  For the symmetrical bandpasses used in Sect.~\ref{sect:filtershapes} we set $c=a$.}
\end{figure}
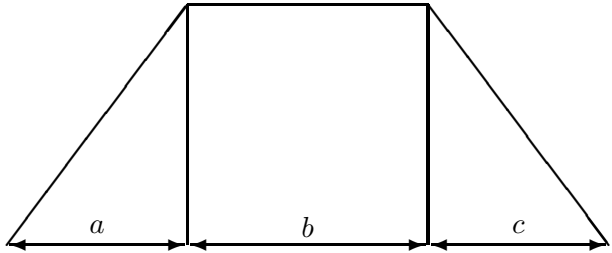

\begin{figure}[t]\begin{center}
\includegraphics[width=84mm]{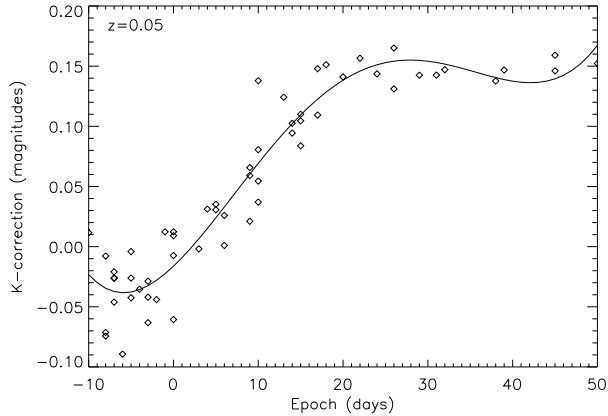}
\caption{\footnotesize K-correction as a function of epoch between -10 and 50 days about maximum for a variety of observed SN Ia spectra all shifted to $z=0.05$, as measured through a trapezoidal bandpass centred on 4500\AA ~with a slope $a=c=500$\AA ~on each side and a flat top $b=1000$\AA ~(integral flux $= 1500$).  We take the RMS over epoch to provide the RMS K-correction value, $K_{\rm RMS}$, for this bandpass (providing one point in Fig.~\ref{fig:Trap2D_Kcorr}).  We take the dispersion about the polynomial fit shown here to provide a measure of the scatter in the K-correction, $\Delta K_{\rm RMS}$, due to intrinsic differences between SNe Ia (one point in Fig.~\ref{fig:Trap2D_RMS}). }
\label{fig:Kcorr-epoch}
\end{center}\end{figure}

\begin{figure}\begin{center}
\includegraphics[width=84mm]{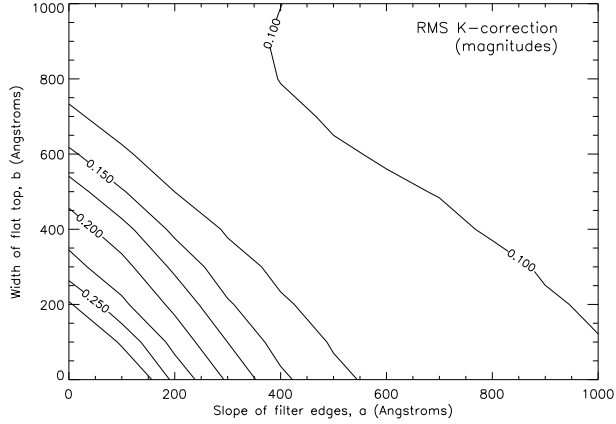}
\caption{\footnotesize K-correction, $K_{\rm RMS}$, as a function of the slope of the wings of the trapezoidal shaped bandpass, $a$, and the width of the flat top, $b$.  The integral flux ($a+b$) increases diagonally from bottom left to top right.  We calculated K-correction as a function of bandpass shape centering the bandpasses on five different positions between 4500\AA\ and 5500\AA.  The same trend appears irrespective of the bandpass position and here we plot the average.  In general the wider the bandpass the smaller the K-correction because the wider bandpasses encompass a larger overlap region of the redshifted and rest frame spectra.}
\label{fig:Trap2D_Kcorr}
\end{center}\end{figure}

\begin{figure}\begin{center}
\includegraphics[width=84mm]{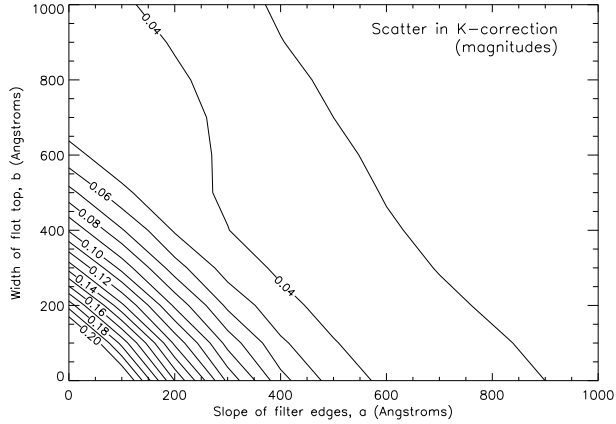}
\caption{\footnotesize We fit a polynomial to the K-correction as a function of epoch and plot the RMS scatter about this fit, $\Delta K_{\rm RMS}$, as a function of bandpass shape.  The shape of the curve is similar to that of the K-correction (Fig.~\ref{fig:Trap2D_Kcorr}), showing that a larger K-correction results in a larger scatter in the K-correction, as expected.}
\label{fig:Trap2D_RMS}
\end{center}\end{figure}\vspace{-5mm}

\begin{figure}\begin{center}
\includegraphics[width=84mm]{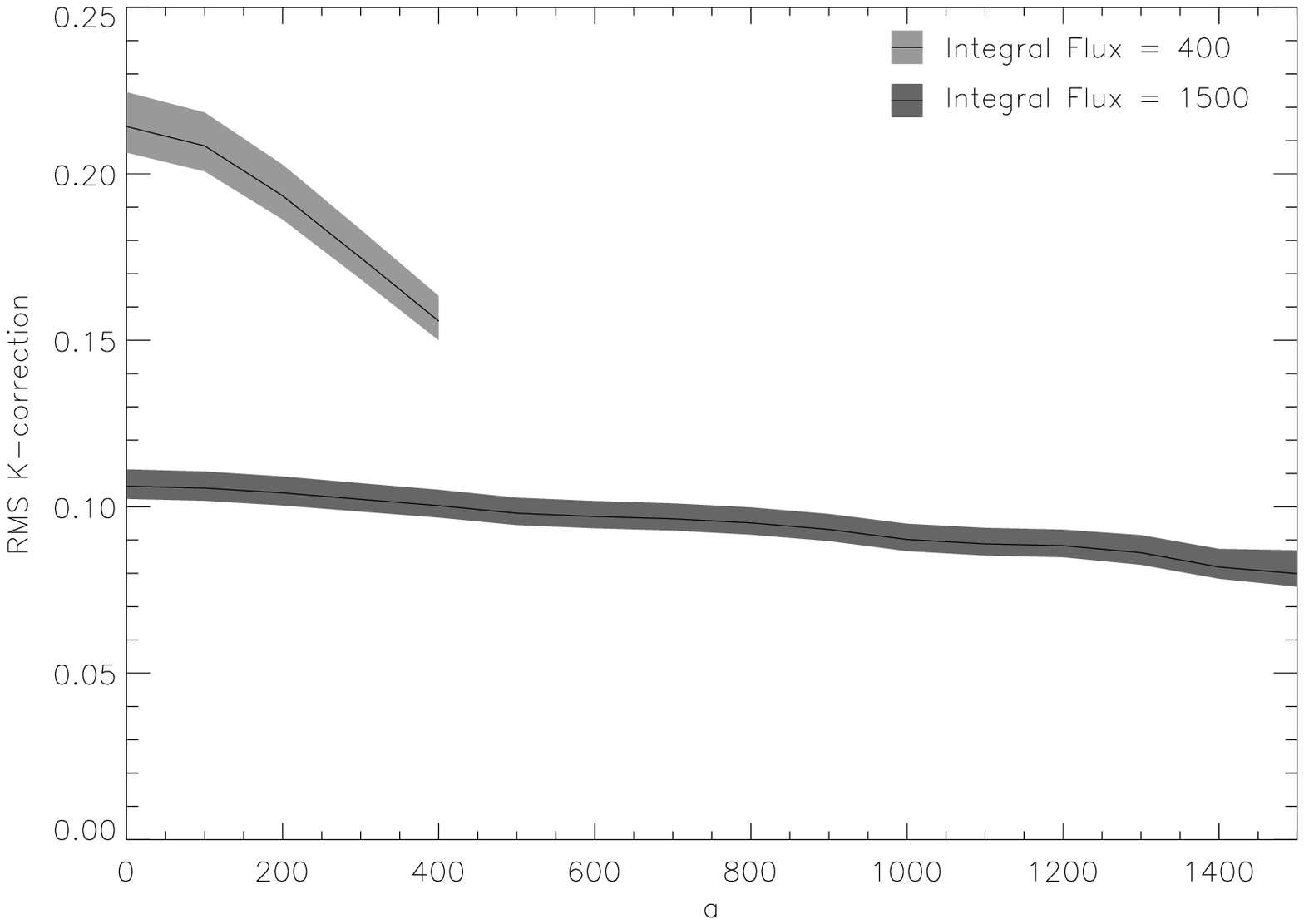}
\includegraphics[width=84mm]{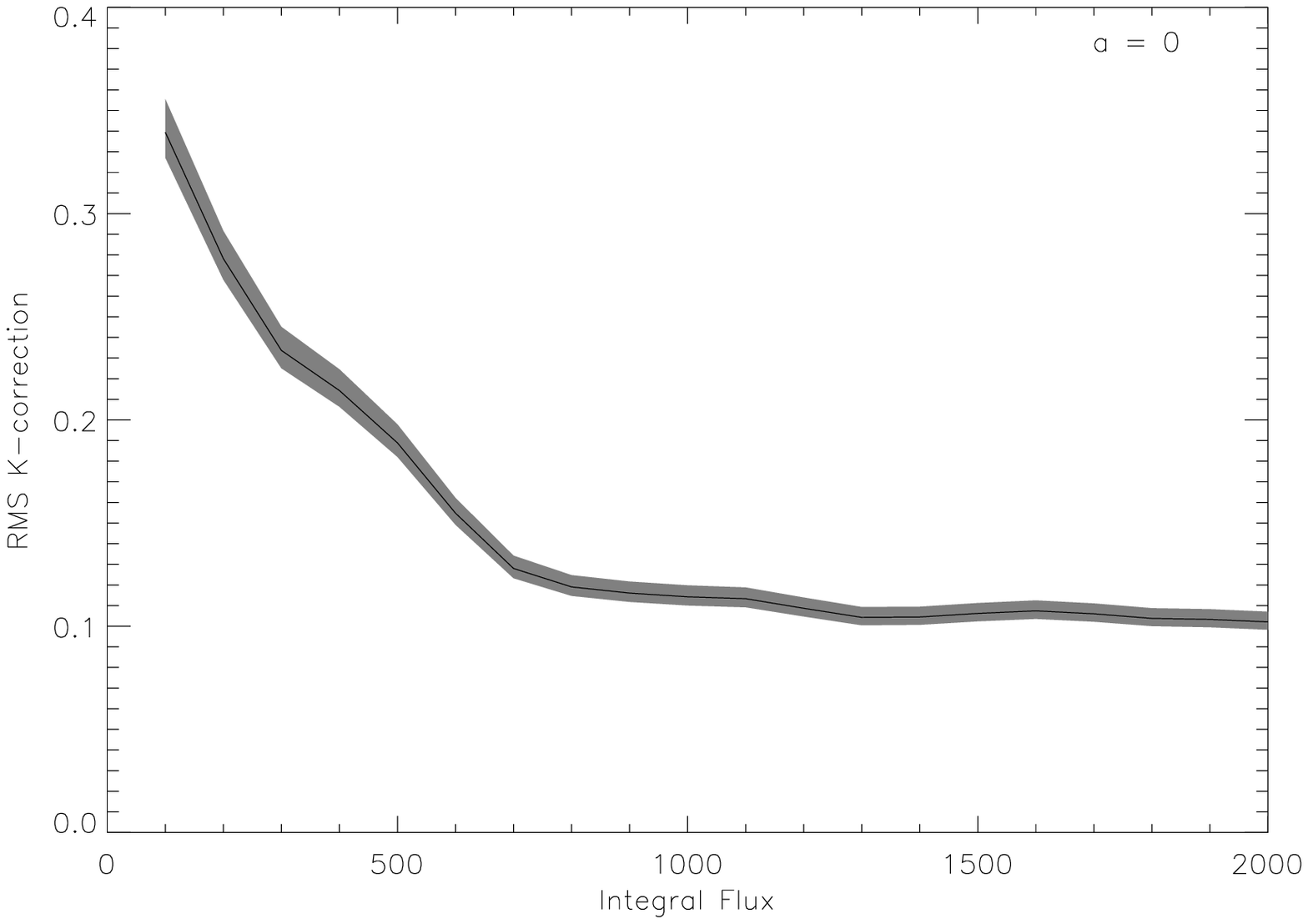}
\caption{\footnotesize K-correction ($K_{\rm RMS}$) vs.\ slope ($a$) and vs.\ Integral Flux ($a+b$).  Edge slope only has a significant effect for narrow filters, the largest improvement comes by increasing integral flux, but as long as the filters are wider than 1000\AA\ little more improvement is gained.  Shaded regions represent 1$\sigma$ uncertainties.} 
\label{fig:Trap2D_RMS_1D}
\end{center}\end{figure}

\vspace{-7mm}
\subsection{K-correction}\label{sect:kcorr}
The K-correction is one of the major systematic corrections that needs to be made on SN Ia observations.  It arises because a given bandpass samples a different spectral region of redshifted and rest-frame supernovae.  The K-correction is the correction required to convert the observed magnitude to a rest-frame magnitude by accounting for the different spectral regions sampled.  We need to perform the K-correction in order to be able to compare the magnitudes of SNe Ia at different redshifts.

Because of the K-correction the most important feature of any bandpass set for SN Ia cosmology is that each bandpass in the set be a redshifted copy of the first.  That is, the bandpasses are the same shape, but redshifted, so that we can sample exactly the same spectral region of high redshift supernovae as we do of low redshift supernovae.  
This reduces the model dependence of the K-correction and thus reduces its susceptibility to systematic error.  Logarithmically spaced bandpasses, where each bandpass differs from the previous one by a multiplicative factor of $(1+z_{\rm SEP})$ for some value $z_{\rm SEP}$, are ideal because they give a pattern of K-corrections that is periodic in redshift, and periodically returns to zero (except for a filter zeropoint offset).  A systematic error could be identified by the periodicity in the results, and corrected because dark-energy models generally do not predict periodic modulation in distance modulus.  The advantage of bandpasses that are self-similar in redshift is so great that we assume this as a required property of any purpose-built filter set for SN Ia cosmology studies.

Using our idealised bandpass shapes we take a sample of 62 type Ia supernova spectra and calculate the K-correction \citep{germany04,nugent02} required to compare their rest frame magnitude to the magnitude that would be observed if they were at a redshift of $\Delta z=0.05$.\footnote{The exact choice of $\Delta z$ is arbitrary --- K-corrections get larger for higher $\Delta z$ but this behaviour is common to all bandpasses so taking a snapshot at a particular redshift is sufficient to quantify their relative behaviour.  Note that because we are using a bandpass set that is self-similar in redshift this $\Delta z$ can be measured from the redshift of any bandpass in the set.  For example if each bandpass is redshifted by $z_{\rm sep}=0.16$ from the previous bandpass then successive bandpasses are at redshifts $z =$ 0, 0.16, 0.35, 0.56... and $\Delta z=0.05$ refers to SNe with $z_{\rm SN} =$ 0.05, 0.22, 0.41, 0.64...  The relationship between these quantities is $1+z_{\rm SN} = (1+z_{\rm sep})^{n-1}(1+\Delta z)$ where n is an integer starting from 1 numbering each bandpass in the set.}    The spectra can be found in  \citet{kirshner75,leibundgut91,filippenko92,phillips92,kirshner93,wells94,patat96,filippenko97} and \citet{reiss99}.  

The K-correction varies according to the epoch of the supernova (Fig.~\ref{fig:Kcorr-epoch}).  The spectra we use cover a range of epochs from 10 days before maximum to 50 days after maximum.  To give a measure of the magnitude of the K-correction for a particular bandpass we take the root-mean-square (RMS) of the K-correction over this epoch range, $K_{\rm RMS}=\sqrt{\langle K(t)^2\rangle}$.  In Fig.~\ref{fig:Trap2D_Kcorr} we plot $K_{\rm RMS}$ as a function of bandpass shape, varying the slope of the bandpass edges, $a$, and the width of the flat top, $b$.  The integral flux through the bandpass ($a+b$) increases along the diagonal.  For each bandpass shape we centre the bandpass at five different wavelength positions between 4500\AA\ and 5500\AA\ so they sample different regions of the spectrum.  The K-correction as a function of shape follows a similar trend irrespective of the central wavelength of the bandpass and the K-correction shown is the average over these different positions.  

In order to measure the scatter in K-correction we fit a polynomial to the K-correction as a function of epoch (
Fig.~\ref{fig:Kcorr-epoch}) and calculate the RMS deviation from this fit, $\Delta K_{\rm RMS}$, (Fig.~\ref{fig:Trap2D_RMS}).   We chose a fourth order polynomial for the fit, which is the lowest order polynomial that fits the data such that there is no residual structure visible by eye.   

The dominant factor determining both $K_{\rm RMS}$ and $\Delta K_{\rm RMS}$ is integral flux, $a+b$.  Wide bandpasses result in lower K-corrections and less K-correction scatter than narrow bandpasses.  This is expected since wider bandpasses naturally have a greater overlap in the spectral region of the redshifted and rest-frame supernovae they sample, so differences in the continuum slope (i.e. the colour) of type Ia supernovae at a particular epoch \citep{wang05}, will have a smaller impact.  Moreover, a variation in a small feature of the spectrum will have less impact on broadband photometry when a wide wavelength range is sampled.  

Plotting the RMS K-correction versus edge slope, $a$, for a fixed integral flux (Fig.~\ref{fig:Trap2D_RMS_1D}) shows there is a slight reduction in the K-correction when we use bandpasses with sloping edges.  This is also expected, since slight velocity shifts, as often seen in SN Ia~\citep[e.g.][]{branch87,benetti05}, can move a supernova feature completely out of the passband of a top-hat bandpass, whereas the same shift would just slightly change the amplitude of the feature in a bandpass with sloping edges.   This is particularly significant for those bandpass and SN redshift combinations that conspire to place a spectral peak on the edge of the passband.  However, integral flux is the dominant factor, sloping edges should not be added at the expense of integral flux.  

K-correction uncertainty improves rapidly up to an integral flux of 1000 (equivalent to the flux under a 1000\AA\ wide top hat), and we  therefore recommend a minimum preferred integral flux of this amount.


\begin{figure}\begin{center}
\includegraphics[width=85mm]{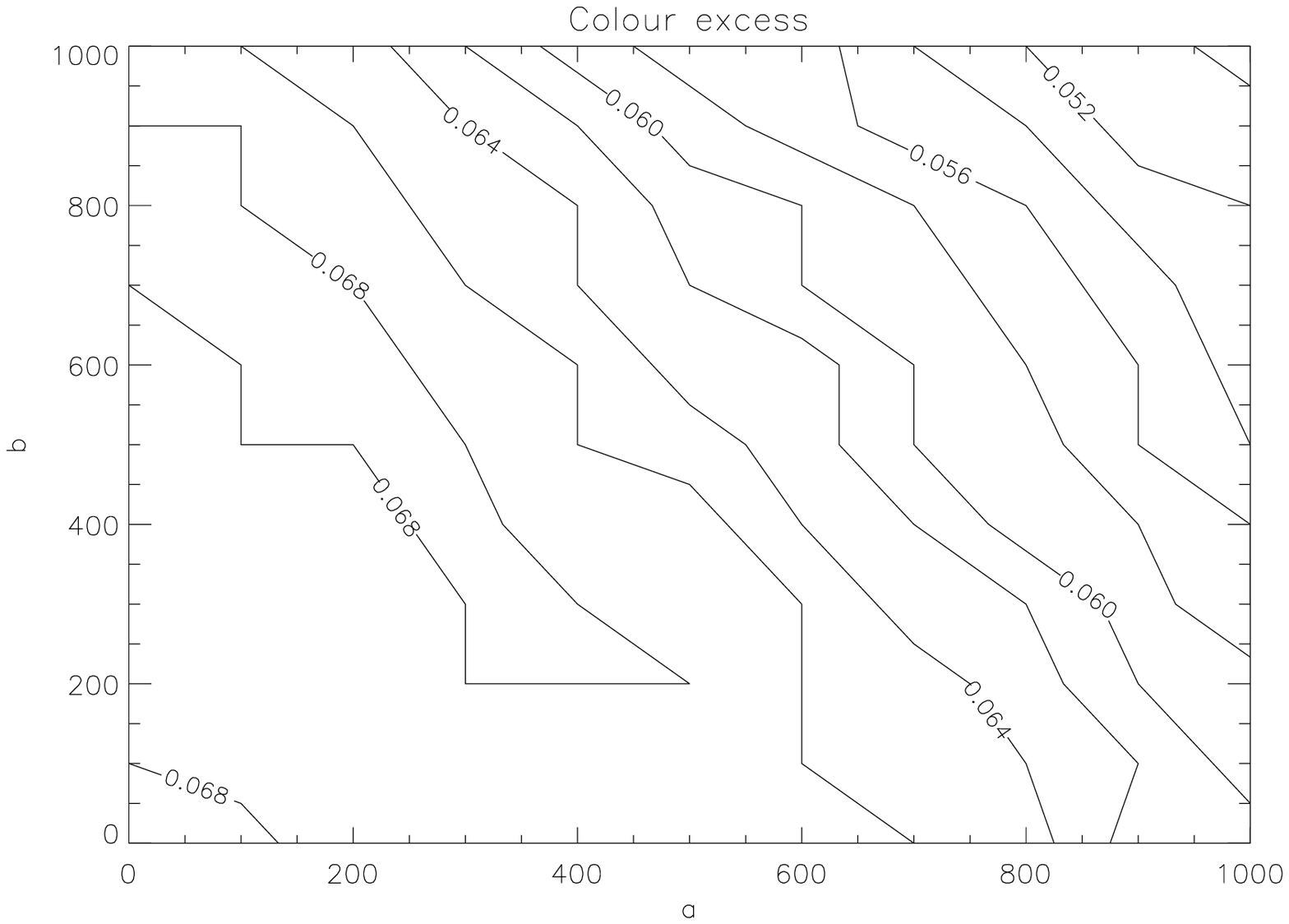}
\caption{\footnotesize Colour excess as a function of bandpass shape, parametrised by the width of the flat top, $b$,  and the width of the sloping wings, $a$, as per Fig.~\ref{fig:Trap2D_Kcorr}.   Shape is less important than the integral flux.  Narrower bandpasses give larger colour excess and thus better colour discrimination and extinction measurement.  However this is a small effect compared to bandpass separation and is also tempered by the increasing K-correction errors as bandpasses narrow. For this figure bandpasses are separated by $z_{\rm SEP}=0.16$.}
\label{fig:Ext2Dshape}
\end{center}\end{figure}

\begin{figure}\begin{center}
\includegraphics[width=85mm]{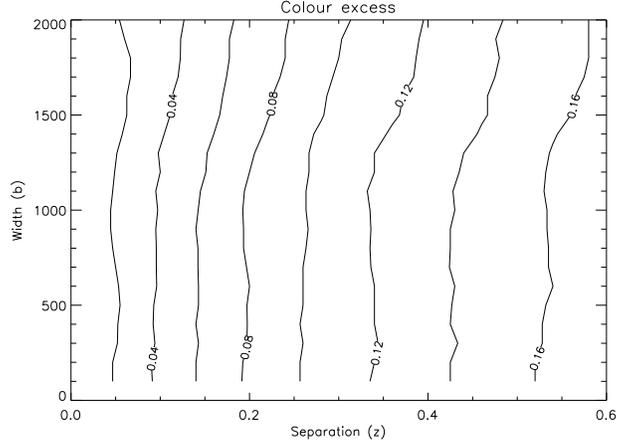}
\caption{\footnotesize Colour excess as a function of bandpass separation and bandpass width.  Bandpasses are all top hats of width b, therefore width is equivalent to integral flux.  Any effect of bandpass width is dwarfed by the effect of bandpass separation.  Widely separated filters are much better at measuring colour excess than closely spaced filters.}
\label{fig:Ext2Dseparation}
\end{center}\end{figure}


\subsection{Extinction correction}

Another important intrinsic feature affecting observed supernova magnitudes is dust extinction \citep{phillips99}.  Dust extinction manifests itself as a wavelength dependent attenuation of the emitted flux, with shorter wavelengths attenuated more than longer wavelengths.  This wavelength dependence allows us to measure the amount of extinction present by measuring, for example, the $B-V$ colour of the supernova compared to the $B-V$ colour of a supernova with no extinction.  This colour difference is known as the colour excess, $E$($B-V$).  The standard extinction law is given in \citet{savage79} who provide a table of attenuation as a function of wavelength for one magnitude of $E$($B-V$).  Giving this attenuation the symbol $A(\lambda)$ a supernova spectrum affected by extinction from dust in the host galaxy will have a flux of,
\beq F_{\rm ext}(\lambda) = F_0(\lambda/(1+z)) \times 10^{-0.4A(\lambda/(1+z))E(B-V)}, \label{eq:extinction}\eeq
compared to the emitted flux $F_0(\lambda/(1+z))$ for a galaxy at redshift $z$.  The extinction in other galaxies may deviate somewhat from the canonical Milky Way extinction law \citep[][Sect.~2.3]{draine03}, but for our purposes this canonical extinction law is sufficient.

The accepted method used to correct for extinction is to warp the spectrum to match the SN colour (the warping is done a variety of ways, but the same general principles apply).  In order to make this correction accurately we need to be able to measure colour accurately.  In particular we need to be able to clearly detect the colour excess caused by extinction, $E$($B-V$).  We therefore test how accurately we can measure extinction with different bandpasses.

We take a model SN Ia spectrum at maximum light \citep{nugent02} and add the effect of dust extinction according to Eq.~\ref{eq:extinction}.  We perform synthetic photometry on this spectrum through a variety of bandpasses and measure the colour both before and after adding extinction.  The larger the observed colour excess for a fixed amount of extinction, the better the lever-arm for detecting dust.  Therefore the bandpass set that gives the biggest colour difference after extinction is our preferred set for extinction correction.  

Because the measurement of colour requires two bandpasses we have an additional variable to consider, bandpass separation.  The bandpass separation in the sets we are considering is best described by the redshift between bandpasses, $z_{\rm SEP}$.  Where the transmission of the $n$th bandpass, $T_n(\lambda)$, is related to the transmission of the first bandpass by, $T_n(\lambda)=T_1(\lambda/(1+z_{\rm SEP})^{n-1})$.
As we will see the bandpass shape is of secondary importance to the bandpass separation in determining colour discrimination.  

In Fig.~\ref{fig:Ext2Dshape} we plot the colour excess as a function of bandpass shape.   In this case each bandpass is separated by a redshift of $z_{\rm SEP}=0.16$.  The dominant effect is again due to integral flux, rather than shape, with narrow bandpasses providing better colour discrimination ability than wider bandpasses.

In Fig.~\ref{fig:Ext2Dseparation} we plot the colour excess as a function of integral flux and bandpass separation.  Although the previous figure showed integral flux makes a more important contribution than shape, its contribution is dwarfed by the effect of separation.  More widely separated bandpasses give a marked improvement in colour discrimination ability over closely spaced bandpasses.  
This trend has a limit.  As bandpasses get further apart the maximum K-correction required becomes higher, and the error in the K-correction reduces the quality of colours we measure.  

An alternative, where possible, is to calculate colour based on non-adjacent bandpasses, e.g. first and third bandpasses in a set, allowing a denser bandpass set for K-corrections than is used for colour calculations.  This only fails for the highest redshift SNe being considered for which supernova optical wavelengths are observed by only two observer filters. 

\begin{figure}\begin{center}
\includegraphics[width=85mm]{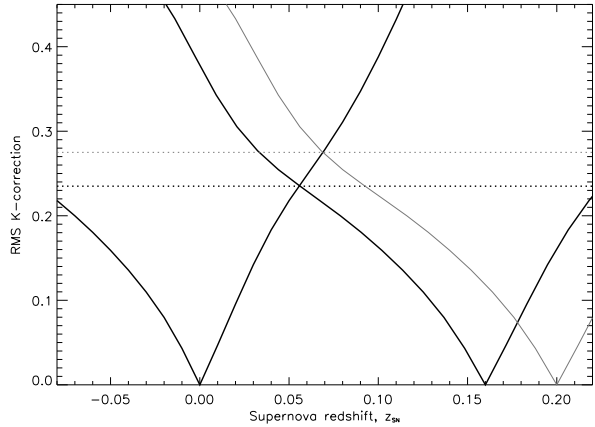}
\caption{\footnotesize $K_{\rm RMS}$ plotted as a function of supernova redshift, for the first two bandpasses in a set.  This diagram assumes the bandpass are separated by $z_{\rm SEP}=0.16$, and uses a bandpass 1200\AA\ wide (approximating the canonical 6-3 bandpass set).  $K_{\rm RMS}$ increases for supernovae redshifted from the centre of a bandpass.  When the redshift becomes large enough it is more economical to consider the supernova blueshifted from the next bandpass in the set.  The maximum K-correction required (dotted line) is approximately midway between these two, and increases when the separation between bandpasses is increased (as demonstrated by the grey curve for $z_{\rm SEP}=0.2$.).}
\label{fig:KcorrZsn}
\end{center}\end{figure}
\begin{figure}\begin{center}
\includegraphics[width=85mm]{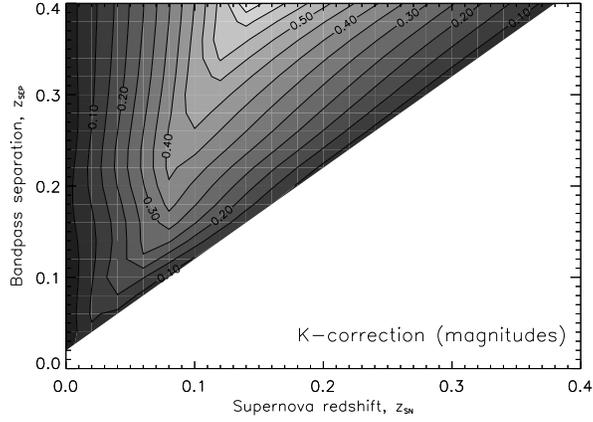}
\caption{\footnotesize K-correction as a function of supernova redshift and bandpass separation.    
Supernova redshift is measured relative to the closest bandpass in the set.  This graph shows only positive redshifts measured from the lower bandpass, but at each point there is a corresponding blueshift from the upper bandpass.  We are interested in colour, so the K-corrections for two adjacent bandpasses are needed (above and below the redshift of the supernova), and here the two values have been added in quadrature.  The K-correction initially increases as the supernovae are redshifted from the lower bandpass, then decreases as the supernovae approach the upper bandpass.  }
\label{fig:KcorrZsepZsn}
\end{center}\end{figure}
\begin{figure}\begin{center}
\includegraphics[width=85mm]{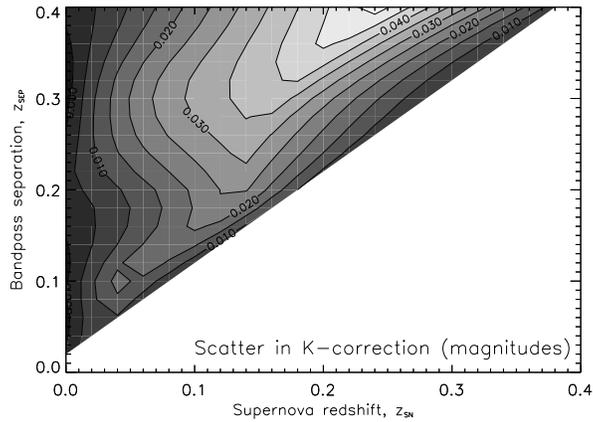}
\caption{\footnotesize Scatter in the colour measurement due to the scatter in the K-correction as a function of supernova redshift (relative to the closest bandpass) and bandpass separation.  We have added the scatter in the two relevant bandpasses (those above and below the redshift of the supernova) in quadrature.  The relative values in this diagram are the key to selecting the best bandpass combination.  The absolute values should be interpreted as upper limits because we expect significant correlation between bandpasses, which would reduce the scatter in the colour.}
\label{fig:ScatZsepZsn}
\end{center}\end{figure}
\begin{figure}\begin{center}
\includegraphics[width=85mm]{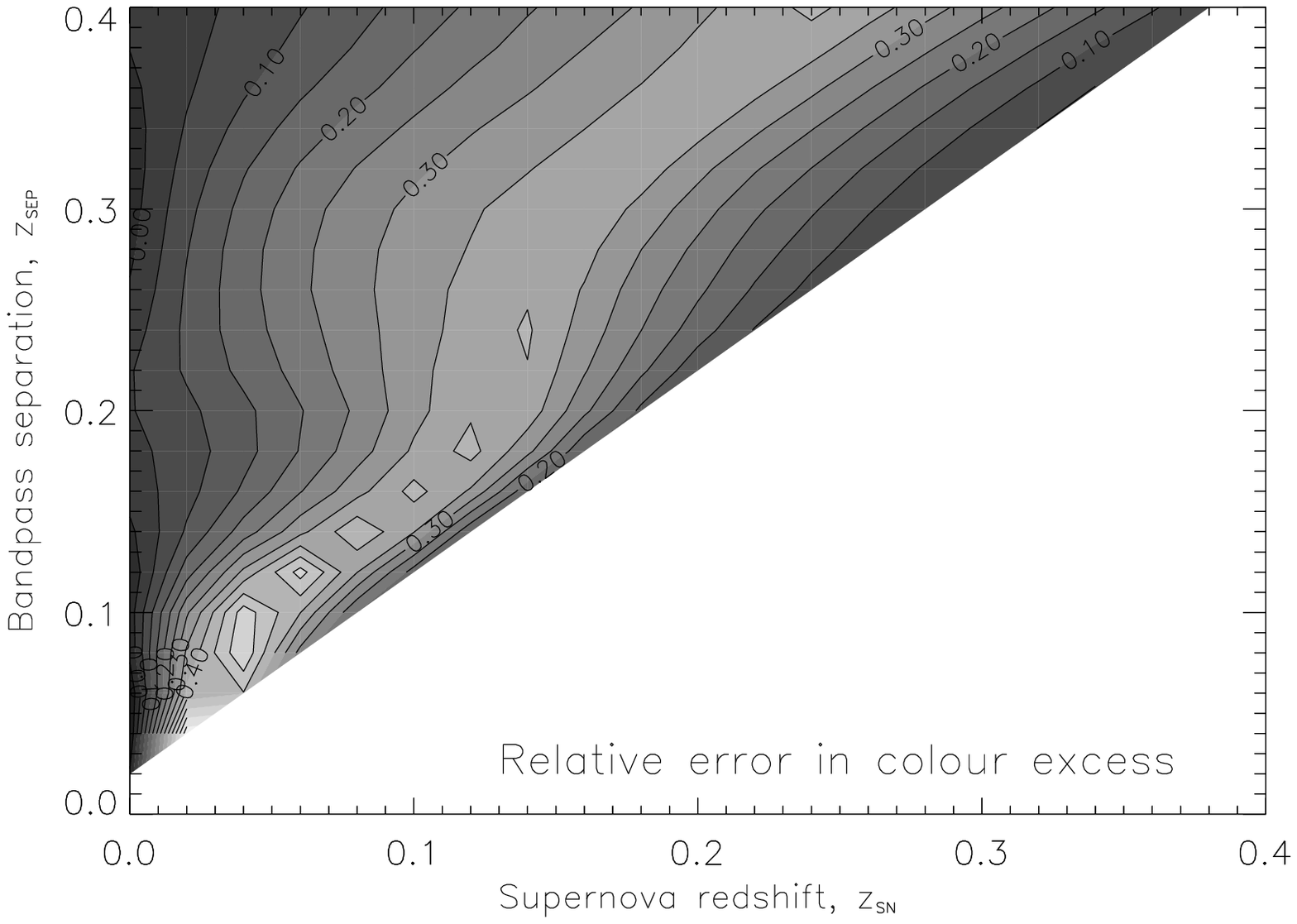}
\caption{\footnotesize Relative error in colour excess as a function of bandpass separation and supernova redshift.  The supernova redshifts are calculated from the nearest bandpass, and this pattern repeats itself up through the bandpass set.  The maximum relative error peaks at low bandpass separations, but is relatively flat overall showing that the improvement in colour measurements through more widely spaced bandpasses is balanced by the degradation of K-corrections.  This can be counteracted by calculating colours using non-adjacent bandpasses.}
\label{fig:ScatEBV}
\end{center}\end{figure}

\subsection{Trade-off}
We have shown that wider bandpasses are better than narrow bandpasses for measuring K-corrections, with a reduction from a K-correction of about 0.3 magnitudes with a scatter of 0.2 magnitudes for a 200\AA\ wide bandpass to a K-correction of about 0.12 magnitudes with a scatter of about 0.04 magnitudes for a 1000\AA\ wide bandpass (Figs.~\ref{fig:Trap2D_Kcorr} and \ref{fig:Trap2D_RMS}, for which $z_{\rm SN}=0.05$).  We have also shown that the width of a bandpass does not have a large effect on measuring colour excess (Fig.~\ref{fig:Ext2Dshape}). Instead, the dominant characteristic for the magnitude of the colour excess is the separation of the bandpasses, with wide separations giving improved colour excess measurements (Fig.~\ref{fig:Ext2Dseparation}).  However, the greater the separation of the filters, the larger the maximum K-correction required (Fig.~\ref{fig:KcorrZsn}).  In general when more than two filters are available colour excess would be measured  using all filters that cover supernova-frame optical wavelengths, not just adjacent filters, allowing for both good K-corrections and good colour excess measurements.  
To investigate whether this gives any strong constraints when only two bandpasses are available (e.g. for the highest redshift supernovae in a sample) we need to calculate the effect of separation on the K-correction and calculate the trade-off with colour.

K-correction increases as a supernova is redshifted out of the bandpass.  It reaches a maximum about half-way between two bandpasses, beyond which it is better to consider the supernova as blueshifted out of the higher bandpass than redshifted from the lower bandpass.  Let the separation of the bandpasses be $z_{\rm SEP}$, and the redshift of the supernova be $z_{\rm SN}$.  For a supernova redshifted from the centre of a bandpass by $z_{\rm SN}^r$ there is an equivalent blueshift, $z_{\rm SN}^b$, from the centre of the next bandpass in the set.  These are related by,
\beq 1+z_{\rm SN}^b = (1+z_{\rm SN}^r)/(1+z_{\rm SEP}).\eeq
In Fig.~\ref{fig:KcorrZsepZsn} we show the K-correction required as a function of bandpass separation and the redshift of the supernova.  We only consider supernovae with redshifts between the first and second bandpasses.  Beyond this the pattern repeats itself with $z_{\rm SN}^r$ measured from the second bandpass and $z_{\rm SN}^b$ measured from the third bandpass.  

The scatter in K-correction for the first and second filters added in quadrature (Fig.~\ref{fig:ScatZsepZsn}) gives the magnitude of the expected scatter in colour due to intrinsic differences in type Ia supernovae.  As our final diagnostic we calculate the relative error this represents in colour excess.  This is plotted in Fig.~\ref{fig:ScatEBV}.  This shows that the error in the K-correction swamps the measurement of colour excess for filter separations below about $z_{\rm SEP}=0.13$.  As separations increase beyond $z_{\rm SEP}=0.13$ the increased error in K-correction balances the increased colour discrimination, so there is no further advantage in increasing the bandpass separation.  In practise, for filter sets with smaller separations the colour excess would be calculated using the first and third (or higher) bandpasses in the set.  This is only a disadvantage, therefore, for the highest redshift supernovae in the sample for which higher bandpasses are not available.

When multiple bandpass measurements are available for each supernova the best lever-arm for extinction is achieved by having many bandpasses over a large wavelength range.  The wavelength range is constrained by the redshifts of the SNe we would like to measure and the wavelengths over which SNe Ia have significant emission, as well as any technological limitations.  Both K-corrections and extinction corrections argue for many bandpasses in a set.  Finite available exposure times argue for few bandpasses.  Ideally the full bandpass set for a type Ia supernova experiment will have as few bandpasses as possible while still keeping the propagated K-correction error (including $E$($B-V$)) less than the required intrinsic dispersion.

Our overall conclusion is that the bandpass pattern should be determined primarily by the need for bandpasses at least 1000\AA\ wide with the widest wavelength range achieveable.
 In Section~\ref{sect:canonical} we design a set of canonical bandpass sets based on the shape and separation considerations we have discussed here and examine the constraints on bandpass separation imposed by detector sensitivities and mirror reflectivity.  

\subsection{Deviations from self-similarity}
Deviations from our assumption of self-similarity of the bandpasses in a set are bound to occur in practice.   
When calculating the effect of deviations from the desired template there are two aspects to consider.  Firstly, if we are unable to accurately measure the deviation in the bandpass we have a calibration error.  We test various types of miscalibration in Section~\ref{sect:calibration} by taking an ideal model spectrum and calculating the error in K-correction the miscalibration produces.  The second aspect occurs when we {\em are} able to accurately measure the deviated bandpass shape.  Error is then introduced because we do not know the exact shape of the underlying SN spectrum, which prevents us from perfectly correcting for the wavelength-dependent variation in the bandpass (this was, of course, the reason we chose self-similar bandpasses).   

The error introduced by this second aspect is sensitively dependent on the shape of the deviation - a variable that is very difficult to quantify due to the multitude of arbitrary shape changes that are possible.  Nevertheless some general principles apply. 
If a bandpass is warped so that it allows greater flux throughput at a wavelength at which type Ia supernovae are particularly variable, then the scatter in the K-correction will increase.  Where that wavelength lies is redshift dependent, so a bandpass that causes a large scatter at $z=0$ will not necessarily also cause a large scatter at $z=0.1$.  Better knowledge of type Ia diversity would allow special consideration of bandpasses with warps in danger regions.
If the bandpass is simply shifted in wavelength from its desired position, but doesn't change shape, then it remains as easy to calibrate as if it was in the correct position.  However, the maximum gap between bandpasses becomes larger and there will be a redshift region (approximately in the middle of this gap) that has a higher scatter than the rest.  If necessary it would be possible to neglect points in this region.

 In general deviations from self-similarity of the bandpasses in a set need to be considered on a case-by-case basis after filters have been produced.

{\small
\begin{table}\begin{center}
\begin{tabular}{|c|ccc|ccc|}\hline
Pattern & \multicolumn{3}{c|}{Range maximised} &\multicolumn{3}{c|}{Usage maximised, $z>1.8$} \\ 
              & $z_{\rm SEP}$ & Width (\AA) & Max $z$&  $z_{\rm SEP}$ & Width (\AA) & Max $z$\\ \hline
 2-2 & 0.469 &  1741 &  1.16&    ---&   ---& ---\\
 3-2 & 0.360 &  1350 &  1.52 &   --- &   --- & ---\\
 3-3 & 0.292 &  1105 &  1.79&  0.294 &  1093. &   1.80\\
 4-2 & 0.288 &  1068 &  1.75&  --- &   --- & ---\\
 4-3 & 0.241 &  1044 &  1.94&  0.229 &  1341. &   1.80\\
 4-4 & 0.203 &  1045 &  2.04&  0.187 &  1513. &   1.80\\
 5-3 & 0.203 &  1045 &  2.04&  0.187 &  1433. &   1.80\\
 5-4 & 0.176 &  1045 &  2.11& 0.158 &  1640. &   1.80\\
 5-5 & 0.155 &  1046 &  2.16& 0.146 &  1406. &   1.96\\
 6-3 & 0.166 &  1023 &  1.94&  0.158 &  1192. &   1.80\\
 6-4 & 0.155 &  1046 &  2.16 & 0.137 &  1654. &   1.80\\
 6-5 & 0.138 &  1046 &  2.21& 0.121 &  1814. &   1.80\\
 6-6 & 0.125 &  1047 &  2.25 & 0.120 &  1294. &   2.10\\\
 7-4 & 0.137 &  1023 &  2.17 & 0.121 &  1433. &   1.80\\
 7-5 & 0.125 &  1047 &  2.25 &  0.108 &  1791. &   1.80\\
 7-6 & 0.114 &  1047 &  2.28 &  0.102 &  1706. &   1.91\\
 7-7 & 0.105 &  1047 &  2.30 &  0.102 &  1215. &   2.20\\
 8-5 & 0.114 &  1047 &  2.28 &  0.098 &  1593. &   1.80\\
 8-6 & 0.105 &  1047 &  2.30 &  0.090 &  1884. &   1.80\\
 8-7 & 0.097 &  1047 &  2.33 &  0.089 &  1579. &   2.02\\ \hline
 \end{tabular}
\caption{\footnotesize Bandpass separations and widths for bandpass sets with different numbers of optical and NIR bandpasses.  The two numbers in the `pattern' column represent the number of optical and NIR bandpasses respectively.  The `Max $z$' column gives the maximum SN redshift we can measure with this bandpass set.  (We require that supernovae at this maximum redshift be observable in the highest {\em two} filters in the set for comparison with low-redshift supernovae colours.)  The two different major columns represent two different possible choices for each pattern.  The first maximises the range - allowing for comparison with the highest redshift supernovae allowed by the detectors.  The second maximises the usage - essentially maximising the width of each filter (and thus how much they overlap) with the proviso that we must be able to measure magnitudes for supernovae out to redshift 1.8 in two bandpasses comparable to the spectral regions sampled for zero redshift supernovae.  All these bandpass sets start with the lower end of the first filter at 3600\AA.}
\label{tab:canonical}
\end{center}\end{table}
}

\begin{figure}\begin{center}
\includegraphics[width=85mm]{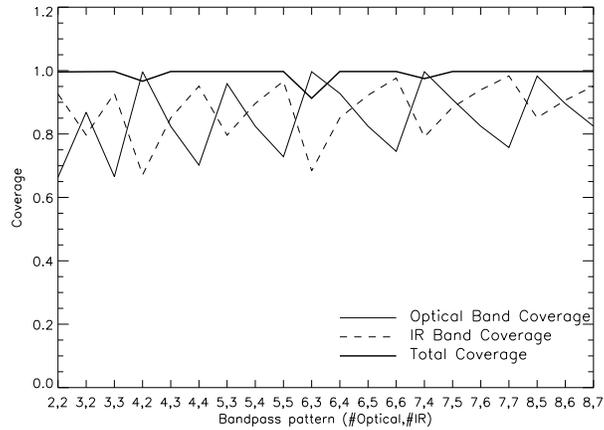}
\caption{\footnotesize Detector coverage for various bandpass sets, where the sets have been chosen so that the maximum wavelength range is covered (Table~\ref{tab:canonical}, columns 2-4).  The bandpasses sets are labelled by the number of bandpasses in the optical passband and the number of bandpasses in the NIR passband.  Coverage is defined as the wavelength range of the bandpass sets divided by the total wavelength range available in the detectors.  There is overlap between the optical and NIR detectors so the total coverage does not equal the sum of the optical and NIR coverage.}
\label{fig:coverage-range}
\end{center}\end{figure}
\begin{figure}[t]\begin{center}
\includegraphics[width=85mm]{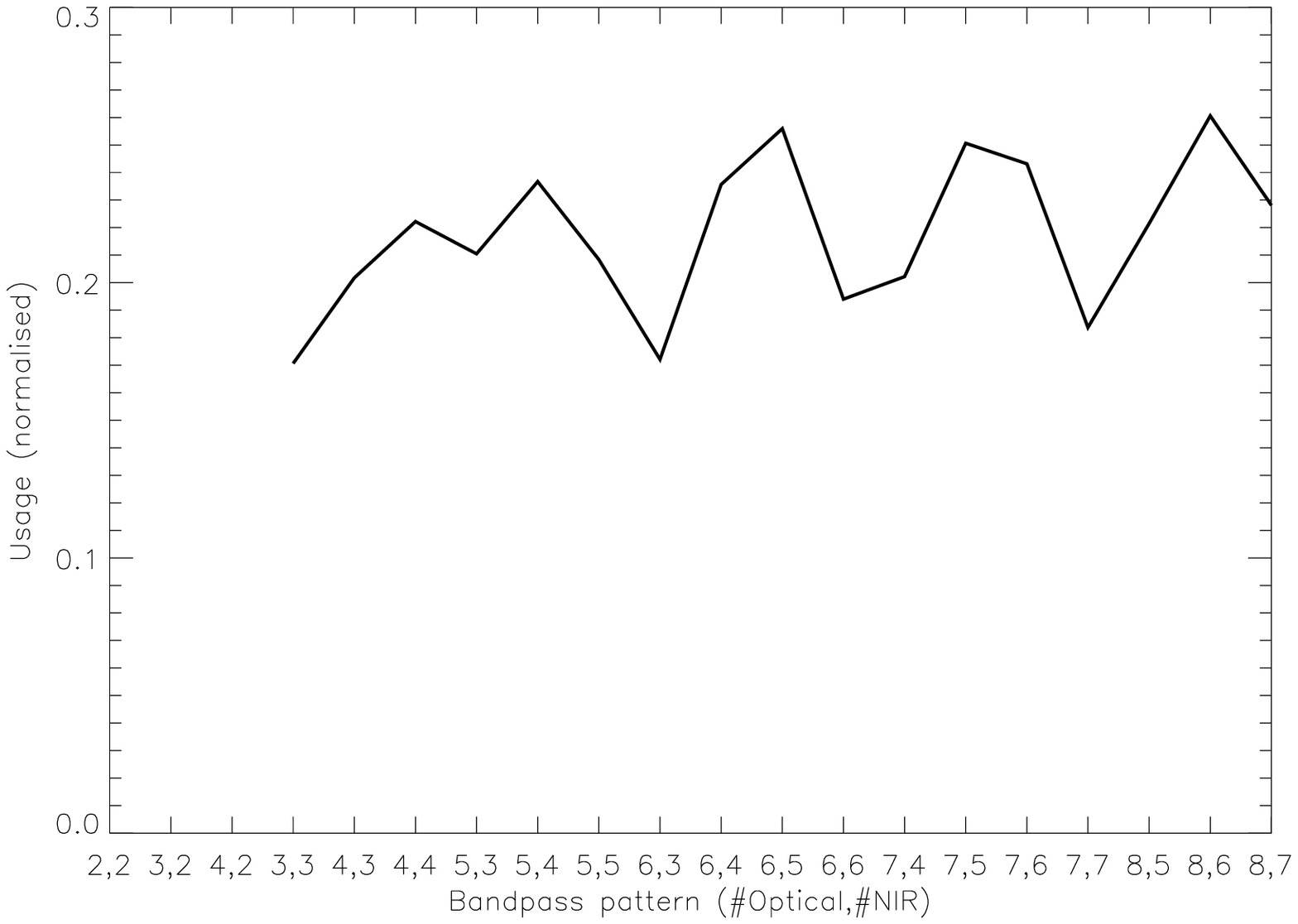}
\caption{\footnotesize `Normalised usage' (as defined in text) for a variety of patterns when usage has been maximised under the proviso that the maximum redshift measurable be 1.8 or more (Table~\ref{tab:canonical}, columns 5-7).  This requirement cannot be satisfied for the 2-2, 3-2 and 4-2 patterns.}
\label{fig:coverage-usage1.7}
\end{center}\end{figure}

\section{Canonical Bandpass Sets}\label{sect:canonical}

In reality we do not have complete freedom in choosing bandpass shapes, and many of the bandpass shapes tested in the previous section are impractical given the detector sensitivities and wavelength coverage we require.  Our primary restrictions are these:
\begin{itemize}
\item{A typical high-resistivity, fully-depleted CCD is sensitive to wavelengths between about 3600\AA  ~and 10000\AA.  (Although sensitive to a little over 10000\AA  ~the detectors may cause focus problems beyond this wavelength because the longer wavelength photons penetrate more deeply into their surface.)}
\item{A typical HgCdTe near-infra-red detector has an adjustable bandgap from 7000\AA ~to 25000\AA~\citep{norton02}.  The example we show in Fig.~\ref{fig:SNspectra-Detectors3} is the detector being considered for the possible next-generation dark energy space telescope, the SuperNova / Acceleration Probe (SNAP), which is a HgCdTe device with a 1.7 micron cutoff.}  
\end{itemize}

Given these restrictions we can make some canonical bandpass shapes and spacings that best use the detector sensitivities.  We design bandpass sets that range from the optical through the near-infra-red to cover the brightest region of a SN Ia from a redshift of zero to a redshift of about $z=1.8$.  This is the most interesting range of redshifts for cosmological parameter estimation because it covers the transition from deceleration to acceleration in the $\Lambda$CMB concordance model.  We have designed these bandpass sets for a space based mission and so are not constrained by the atmospheric emission and absorption that limit ground-optimized bandpass sets.

The results from Section~\ref{sect:filtershapes} suggest that we should use bandpasses at least 1000\AA\ wide, ideally with some slope on the edges, and have many filters covering the total detector range to keep the required K-corrections low.  When the separation of the bandpasses is kept to less than $z\sim0.22$ the scatter in the K-correction remains below 0.02 magnitudes.\footnote{See Fig.~\ref{fig:ScatZsepZsn}, where the scatter in K-correction for a {\em single} bandpass is approximately $\sqrt{2}$ of the scatter values shown}    
Previous studies~\cite{kim03,kim03b} performed for the SuperNova / Acceleration Probe (SNAP) design have shown that a combination of six optical filters and three infra-red filters can provide K-correction errors of  less than 0.02 magnitudes and is technically feasible.

Let the $n$th bandpass range from $\lambda^{\rm min}_n$ to $\lambda^{\rm max}_n$.  If we call the lowest wavelength bandpass $n=1$ then the rest of the bandpasses are related to the first by,
\beq \lambda_n = (1+z_{\rm SEP})^{n-1}\lambda_1,\eeq
where $z_{\rm SEP}$ is the redshift spacing between the bandpasses.  If the width of the first bandpass is $\Delta \lambda$ then $\lambda^{\rm max}_n = \lambda^{\rm min}_n + (1+z_{\rm SEP})\Delta \lambda$.  To fit in the range of the typical detectors we are considering the optical bandpasses need to fit in the range 3600\AA-10000\AA, while the NIR bandpasses need to fit in the range 7100\AA-17000\AA.  
Adding the requirement that the bandpasses must overlap, means $\lambda^{\rm min}_1 z_{\rm SEP} < \Delta \lambda.$

These restrictions leave a very limited range of bandpass sizes and spacings that will fit within the detectors.  We have free rein, however, to choose the number of bandpasses to put in each detector.   We create a series of bandpass sets with various patterns -- i.e.\ various numbers of bandpasses in the optical and NIR.  For each pattern we can choose to optimise the width and spacing of the bandpasses a number of ways.  One way is to maximise the range covered -- so we can measure the highest redshift supernovae allowed by the detectors.  Another way is to maximise the usage of the bandpass - requiring the maximum overlap of the bandpasses so the flux is greatest and the survey can be done most quickly.  In practise a balance between these two is likely to be most useful.  In Table~\ref{tab:canonical} we provide the details of bandpass sets optimised firstly for maximum range and secondly 
for maximum usage with the additional proviso that they must be able to observe supernovae out to at least a redshift of 1.8 -- the approximate range of sensitivity expected for the next generation of space telescopes -- in two bandpasses.  

We define `coverage'  to be the wavelength range covered by a bandpass set compared to the range available in the detectors.  Figure~\ref{fig:coverage-range} shows the coverage for a variety of bandpass sets, where for each pattern we have chosen the set that maximises the wavelength range covered (Table~\ref{tab:canonical}, columns 2-4).  Most patterns are able to be designed to completely cover the detector wavelength region.  An exception is the 6-3 pattern, which is too highly constrained by fitting 6 bandpasses in the optical region to be able to cover the IR region with just 3 bandpasses.  Similar problems occur in other bandpass sets that have unbalanced numbers of optical and NIR bandpasses.

We define `usage'  to be the sum of the bandpass widths in a set divided by the total wavelength range of the detectors.  To take into account the greater survey time required for a larger bandpass set we also divide by the number of filters in the set to get a `normalised usage'.   Figure~\ref{fig:coverage-usage1.7} shows the normalised usage for various patterns when the bandpass sets have been chosen to maximise the usage while still being able to observe supernovae out to $z=1.8$ (Table~\ref{tab:canonical}, columns 5-7).  The patterns 2-2, 3-2 and 4-2 cannot satisfy these criteria.  Bandpass sets with slightly fewer NIR  than optical bandpasses can achieve the best signal-to-noise in the shortest time.  

The detectors do not have good performance near the cut-off points, so the edges of the bandpasses need to be tuned to match the edge of the detector performance if we want to utilise as much of the detector sensitivity as possible.  The numbers given here are a rough outline for recommended widths and spacings, the specific shape of the bandpasses is flexible, as long as sloping edges do not reduce the flux through the bandpass below the equivalent of a 1000\AA\ wide top hat.


\section{Calibration constraints}\label{sect:calibration}

Errors in filter calibration affect every measurement made with that filter over the lifetime of the experiment.  In particular, because a different filter is used to measure the photometry in different redshift bins, a systematic error in filter calibration may introduce a spurious redshift dependent trend.  For supernova observations constraining the luminosity distance this is among the worst type of systematic error. 

The two main filter types are glass and interference filters.  Each has advantages and disadvantages.  Interference filters have a high throughput and can be designed to match arbitrary bandpass shapes, even in the infra-red.  However  the bandpasses of interference filters are dependent on the angle-of-incidence of the incoming light and this varies with position on the focal plane.  This can be calibrated-out to some degree by applying, for example, a position dependent K-correction \citep{kim03,kim03b}, or alternatively using the bandpass for the average incidence angle and allowing the position dependence to contribute to the scatter about the mean.  This effect contributes an error on the order of 0.01mag \citep{kim03b}.  
Experience shows that interference filters are subject to more variability with time than glass filters, in particular outgassing can change the layer thickness by a substantial and irreversible amount.   
They are also more susceptible than glass to radiation damage 
and are difficult to manufacture to an exact and repeatable specification.

Glass filters are more stable than interference filters.  They can be made very uniform over a large area, and very stable with time~\citep{schottglass}.  Their bandpass has negligible angle dependence.  However, they cannot be tuned to arbitrary bandpass shape and they have lower throughput than interference fitlers ($\sim 60\%$ for glass filters as compared to $\sim 80\%$ for interference fitlers).  Most importantly there are no adequate glass filters available in the near-infra-red, where a high redshift SN Ia experiment needs to probe.  

It may be that the most stable and accurate filter set for SN Ia cosmology could be built from a combination of glass and interference filters or possibly  
 hybrid glass / interference filters where glass provides the low wavelength cut-on and a small number of interference filter layers profides the cut-off.  The novel idea of using high pass filters with tuned cut-on points and then subtracting them post-observation to create a narrower filter may also be worth considering.  However, this procedure increases the measurement error. 
 Unless glass filters are developed in the infra-red the NIR filters need to be interference filters.

Experience has shown that in the manufacture of interference filters we can expect a non-trivial difference of $\sim 5\%$ at any particular wavelength with respect to transmittance function specification~\citep{jbh}.  If we can perform calibration perfectly after manufacture this is not an issue.  The difficulties come in (a) positional dependence due to different incident angles when the filters are in a converging or diverging beam (b) miscalibration due to non-uniformity and (c) dependence of transmittance on temperature, humidity and general decay with age.  Transmittance of filters depends on temperature at the level of $\sim 0.2$\AA$/K$~\citep{jbh} and if filters absorb or lose moisture after the calibration procedure it will permanently and significantly alter their bandpasses.  Calibration needs to be done with the filters and detectors installed in an operational telescope, and needs to be done for many positions and many angles.  
  
Miscalibration of filters is less important if the miscalibration is sufficiently random such that the mean transmittance can be accurately determined with the errors caused by the various effects gaussianly distributed about the measured mean.  The error would then contribute to the random uncertainty and be accessible to improvement with better statistics (more SNe).  Problems arise, however, if the calibration technique tends to err in one direction for all filters, or if environmental/age related effects push all filters in a common miscalibration direction.

It is important, therefore, to consider the accuracy of calibration in the choice of a filter set for future SN Ia studies.  There are many different filter types and manufacturing techniques, and it is beyond the scope of this paper to examine each in detail.  We provide instead 
the calibration accuracy needed (noise amplitude) in order to reach a particular photometric accuracy requirement.

\begin{figure}\begin{center}
\includegraphics[width=88mm]{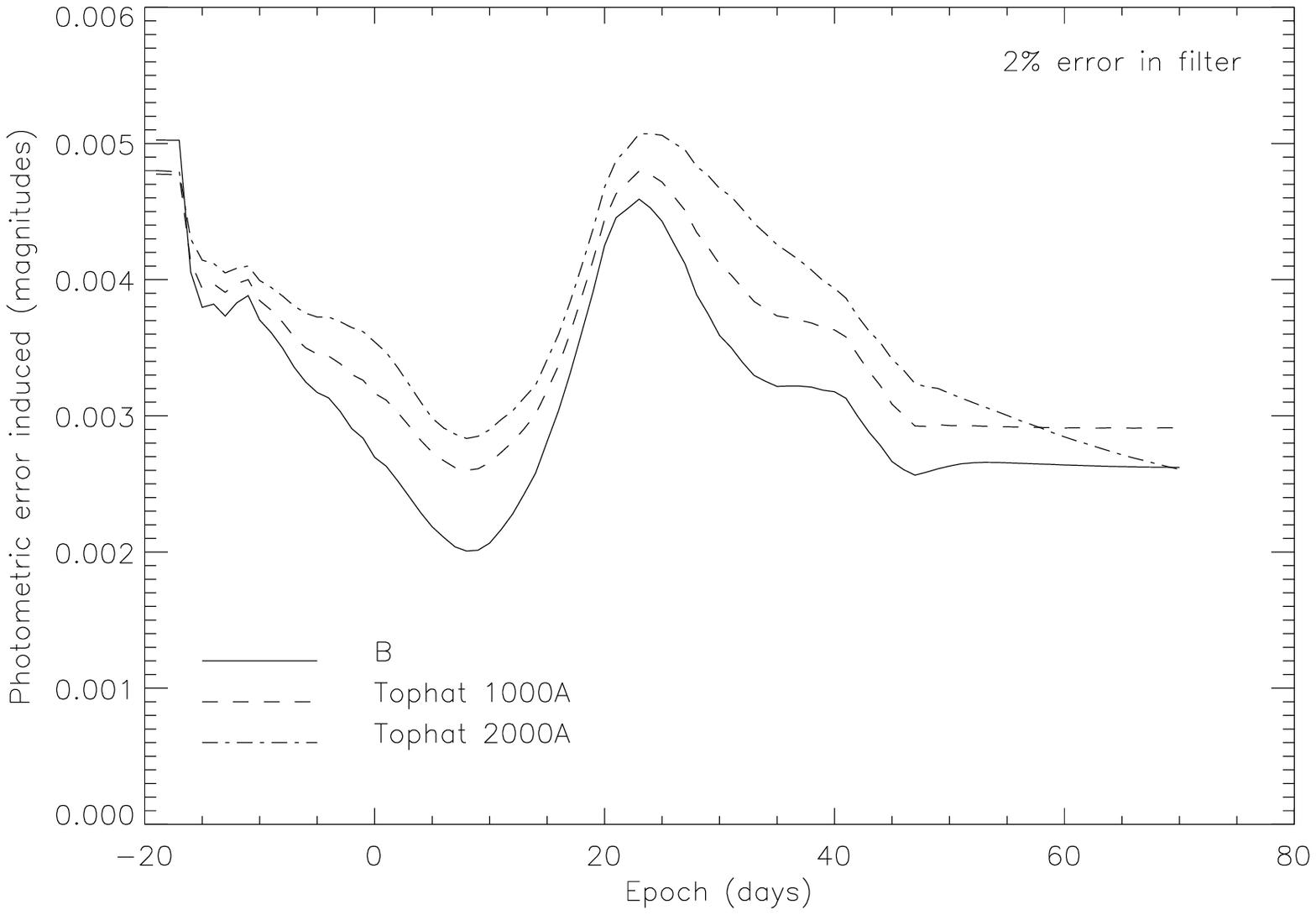}
\caption{\footnotesize Photometric error as a function of supernova epoch for noise amplitude levels of 2\% applied to three different filters.  The first is a Bessell B filter.  The other two are top hat filters, centered on 4500\AA\ with widths of 1000\AA\ and 2000\AA.  The error induced is independent of the filter shape and instead dependent only on the amplitude of noise introduced.  Error increases slightly as the integral flux through the filter increases.}
\label{fig:Phot-Epoch}
\end{center}\end{figure}

\begin{figure}\begin{center}
\includegraphics[width=88mm]{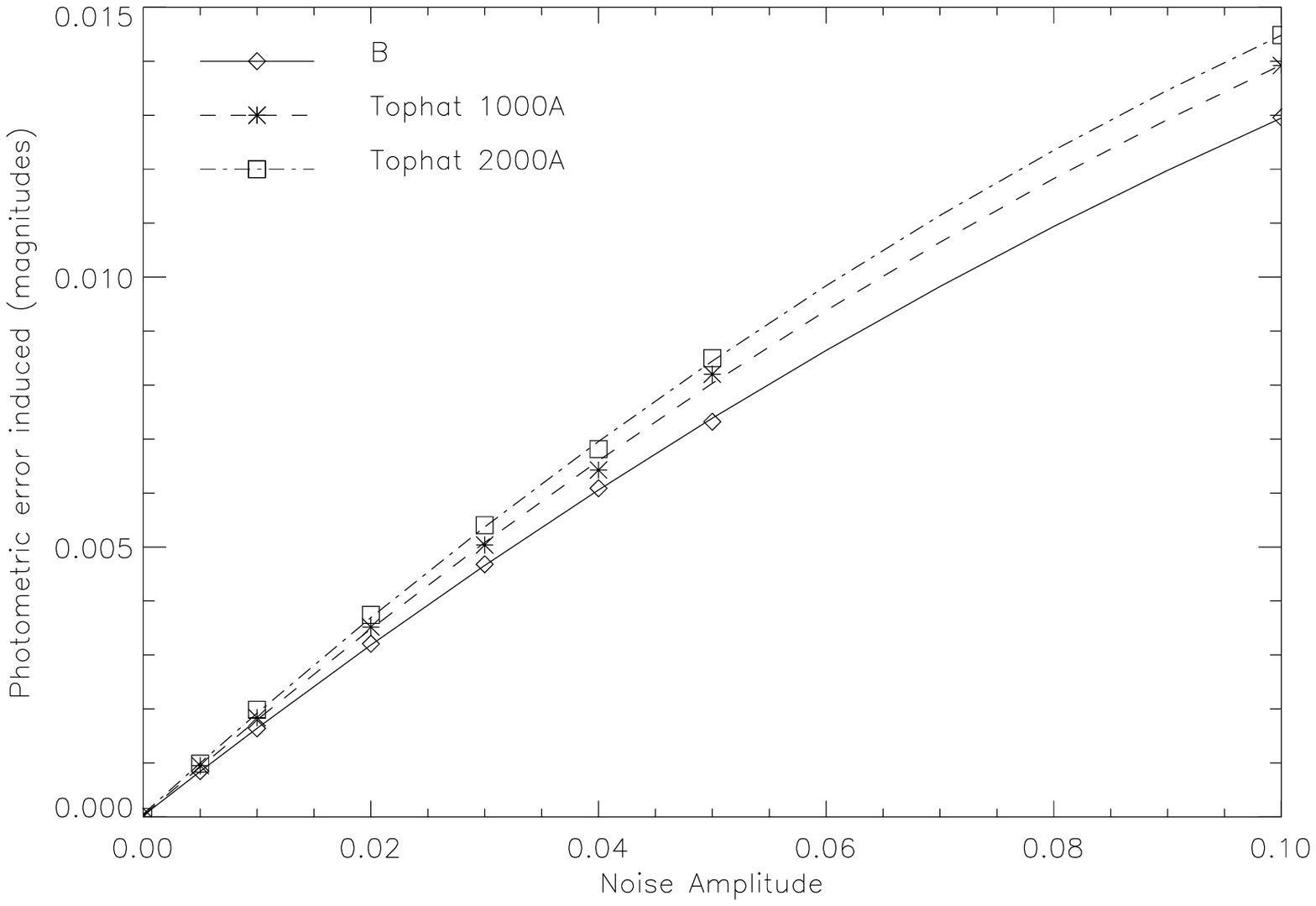}
\caption{\footnotesize Comparison of average error in photometry introduced by uncharacterised manufacturing error in a Bessell B filter, and two top hat filters of different widths (1000\AA\ and 2000\AA).  For each noise level we test up to 1000 realisations of a noisy filter and calculate the RMS error in photometry as a function of epoch of observation.  We then calculate the RMS over all epochs to determine the final error magnitude vs. noise amplitude plotted here.  The fits are quadratic.}
\label{fig:Phot-Noise}
\end{center}\end{figure}

\subsection{Random mis-calibration}\label{sect:random}
The process of creating a filter involves initially designing an ideal filter that the manufacturers will try to generate.  There will always be some level of manufacturing error in this process, so calibration is necessary after manufacture and again once deployed (e.g.\ calibration must be done in orbit for a space mission).  We assume that this calibration cannot be done perfectly and may also change with time and wear.  
We therefore assess to what level of accuracy the calibration is needed in order to reach photometric accuracy requirements.  Is it worth using interference filters to create specialised bandpasses if glass is more stable and can be calibrated more accurately?

As previously discussed, filters are two dimensional objects that cover the focal plane, so their bandpass may vary with position.
Characterisation of the filter must therefore either specify the bandpass as a function of position, or assume the filter is uniform across its surface to a certain precision.  In the following we define a noise amplitude for which we calculate the corresponding uncertainty in photometry.  To permit the assumption of uniformity the filters must be uniform across their surface to the precision given by the noise amplitude for the required photometric accuracy.

In order to simulate an unknown calibration error in a filter we take a filter template and multiply it by noise with a flat power spectrum.  
We call the filter function measured during the calibration process the `template' filter.  We call the actual filter we have the `noisy' filter.
We then take the library of idealised SN Ia spectra from \citet{nugent02} and calculate the difference in the measured photometry between the template and noisy filters as a function of epoch (days before and after maximum light).   The resulting photometric difference represents the systematic error introduced by the uncharacterised noise in the filter.
%
Each realisation of noise produced has a different profile, and a different scatter.  When calculating tolerances we generate many realisations of a noisy filter and take the RMS of the error.  

White noise is probably not appropriate for all $k$, since a real filter will not have noise at the smallest scales, but these simulations also have a natural lower limit through the finite number of array elements we use to describe the filter.  We varied the density of the array and it does not affect the results.

We performed the above analysis on three different types of filters: the Bessell B filter, a top hat filter 1000\AA\ wide and a top hat filter 2000\AA\ wide.  Both of the top hat filters are centered on 4500\AA.  

The supernova spectra vary as the supernovae evolve, therefore the error in the filters is epoch dependent.  In Figure~\ref{fig:Phot-Epoch} we plot photometric error vs.\ epoch for a noise amplitude level of 2\% in three different filters.  We repeat this analysis for noise amplitude levels from 0.5\% to 5\%, then average over all epochs to come up with a single error specification as a function of noise level.  This is plotted in Figure~\ref{fig:Phot-Noise}.  The shape of the filter profile is not significant in this analysis, only the quantity of noise we are adding.  
The slight differences in the curves arises because the wider filters sample a wider region of the supernova spectrum.

It is usually argued that systematic errors should be added directly, not in quadrature.  So if each filter has a systematic uncertainty of 0.002 magnitudes then the colour has an uncertainty of up to 0.004 magnitudes (correlation between bandpasses reduces this number).  Extinction, $A_{V}$ is given by $A_{V}=3.1\;E({\rm B}-{\rm V})$, so with the above parameters the systematic uncertainty in extinction is up to 0.012 magnitudes.  This already represents much of the error budget for photometry with proposed space missions (e.g. SuperNova / Acceleration Probe, SNAP).

To achieve an irreducible systematic photometric accuracy of better than 0.002 magnitudes in a single filter we need to create filters that are uniform across their surface (or with a known position-transmittance function) and characterised to better than $\sim$1\% (noise amplitude less than about 1\% of peak transmittance).    This criterion gets slightly stricter for filters with larger integrated flux.  

Note that although the average error converges on the curves shown in Fig.~\ref{fig:Phot-Noise}, any one realisation of a noisy filter can have an error that differs significantly from this, and may be up to twice as large. 
This may more accurately represent the situation we are faced with since any filter we build will be just one realisation of a noisy filter.  

We tested the photometric error induced when the supernova spectra are slightly red- or blue-shifted with respect to the filter.  The profile of photometric error vs.\ epoch appears slightly different,  
but the average photometric error is unchanged. 

\begin{figure}[t]\begin{center}
\includegraphics[width=90mm]{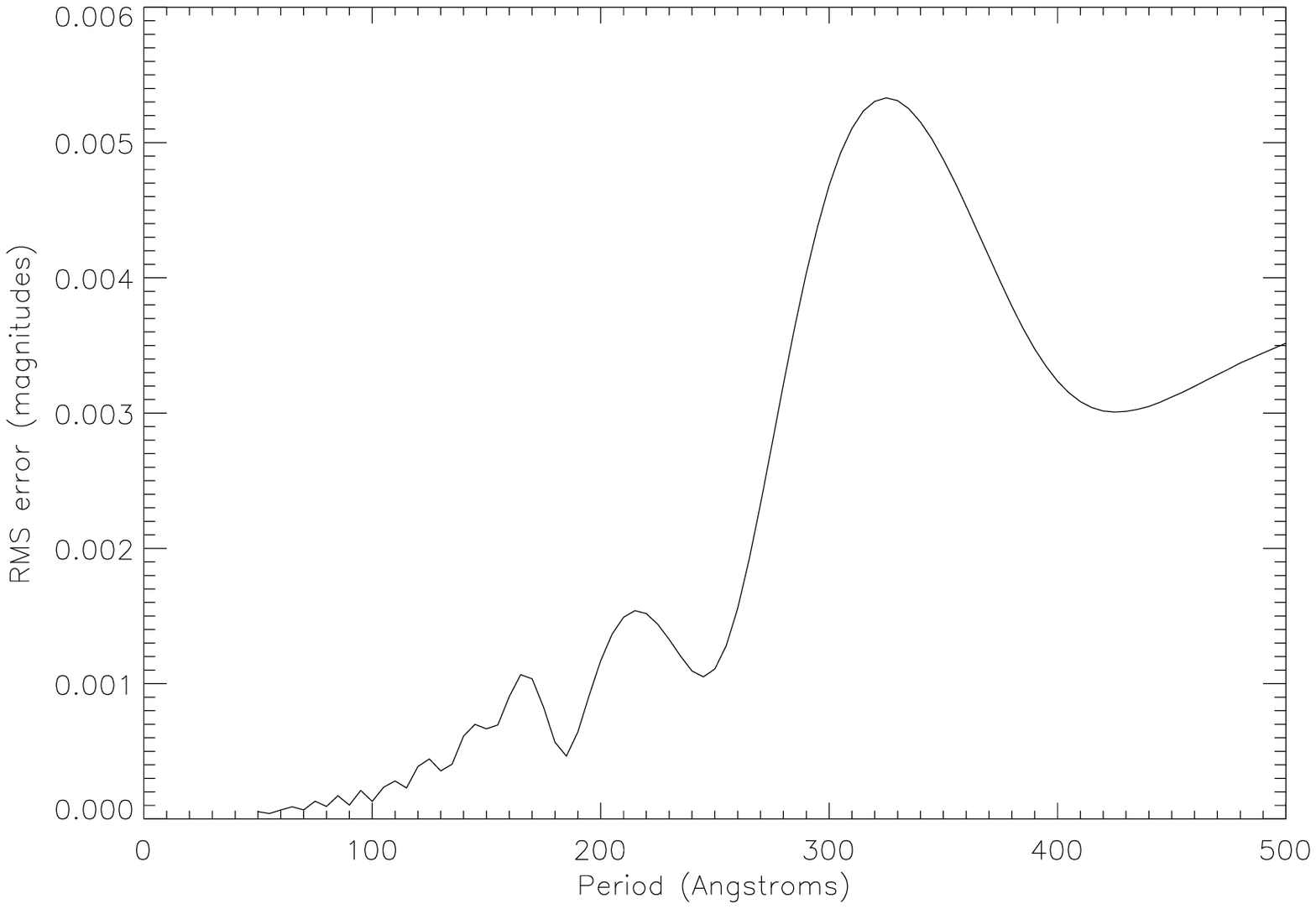}
\caption{\footnotesize How photometric error in measurements of supernovae depends on a periodic uncertainty in filter calibration.  We have averaged over phase shifts from 0 to 2$\pi$.  This indicates that small scale (short period) variations have less impact than large scale (large period) variations, and that any filter that has a periodicity of around $330\pm50$\AA\ should be avoided.}
\label{fig:averageperiod}
\end{center}\end{figure}

\subsection{Periodic mis-calibration}\label{sect:periodic}
With interference filters there is a possibility that any miscalibration may be periodic in wavelength.  We therefore examine the effect of a periodic miscalibration to see whether there are any frequencies that should be avoided for supernova observations.  

We take a template filter and multiply it by a single Fourier component (a sine wave) that represents our unknown periodic calibration error.  We have thus created another type of `noisy' filter.  We create a series of noisy filters by varying the period of the noise that we add to the template, while keeping the amplitude of the noise constant in all cases.  This allows us to calculate the error as a function of SN epoch that we would introduce if we used the noisy filter thinking we had the template.
Taking the RMS of this error gives us a single number quantifying the error introduced as a function of the periodicity of the noise.   For each period we also re-calculate the RMS error for phase shifts from 0 to $2\pi$ and take the average of the result.  The effect of phase is significant due to spectral features moving in and out of peaks and/or troughs of the miscalibrated filter. 

Figure~\ref{fig:averageperiod} shows the effect of period on the RMS error between a template filter and a noisy filter.  The noise amplitude has an RMS of 5\% of the transmittance, and the periods range from 50\AA\ to 500\AA. Over the wavelength range shown here the error increases as the period of the miscalibration gets higher.  
This arises because type Ia supernovae spectra are relatively smooth on small scales ($\sim50$\AA) but have significant features on larger scales (several hundred Angstroms).  When the periodicity of the miscalibration in the filter matches the dominant frequency in the power spectrum of the supernova, the error due to miscalibration is largest. This occurs around 330\AA.
From this (Fig.~\ref{fig:averageperiod}) we conclude that any periodicity of around $330\pm50$\AA\ should be avoided in the manufacture of filters for SN Ia studies.

\section{Conclusions}
K-correction scatter due to intrinsic diversity in the SN Ia population can be minimised by using bandpasses wider than 1000\AA.  Adding sloping edges also helps, particularly in the few cases where a spectral line sits on the edge of a bandpass, but should not be added at the expense of integral flux.  Extinction corrections rely on the accurate measurement of colour, which is done best using widely spaced filters, irrespective of their width or shape.  However, there is a comprimise because more widely spaced filters mean larger K-corrections, which in turn increase the error in colour calculations.  The trade-off suggests that spacings wider than $z_{\rm SEP}=0.13$ should be used, but this is not necessarily a strict lower limit because colour can be calculated using non-adjacent filters.  It is a genuine lower limit if colours and extinction corrections are required for the highest redshift supernovae in the detector passband.  This paper addresses the relation between passband set and intrinsic supernova dispersion in K-correction.  Future work will examine intrinsic dispersion in absolute magnitude.

Bandpass spacing and width are the determining factors for optimally fitting an evenly spaced bandpass set in the detector sensitivities.  We investigated how well different bandpass patterns fill the sensitive regions of the detectors and provide specifications for several optimised bandpass sets in Table~\ref{tab:canonical}.

The photometric accuracy achievable through different kinds of filters is an important aspect of systematic error calculations.  Our analysis here shows that characterising the filters to better than $\sim1$ \%, as defined in Section~\ref{sect:random}, keeps the photometric error in a single filter below 0.002 magnitudes.  This is independent of filter shape but increases slightly for fliters with larger integrated flux.  Periodic error in filter calibration is to be expected for interference filters.  Filters that may cause a periodic error at around $330\pm50$\AA\ should be avoided.\vspace{-1mm}

\acknowledgements
TMD and BPS acknowledge Australian Research Council Grants LX0454445, DP0209028, and DP0559024.
AGK is supported by the Director, Office of Science, of the U.S.\ Department of Energy under Contract No.\ DE-AC02-05CH11231.
TMD and AGK would like to thank the Aspen Centre for Physics for their hospitality during early work on this project.  


\newpage

\end{document}